# A Simple Extension of Dematerialization Theory: Incorporation of Technical Progress and the Rebound Effect

Christopher L. Magee
Massachusetts Institute of Technology, Institute for Data, Systems, and Society,
 77 Massachusetts Ave, Cambridge, Massachusetts 02139
Tessaleno C. Devezas
Faculty of Engineering, University of Beira Interior, 6200-001 Covilhã, Portugal

**Abstract:**
Dematerialization is the reduction in the quantity of materials needed to produce something useful over time. Dematerialization fundamentally derives from ongoing increases in technical performance but it can be counteracted by demand rebound - increases in usage because of increased value (or decreased cost) that also results from increasing technical performance. A major question then is to what extent technological performance improvement can offset and is offsetting continuously increasing economic consumption. This paper contributes to answering this question by offering some simple quantitative extensions to the theory of dematerialization. The paper then empirically examines the materials consumption trends as well as cost trends for a large set of materials and a few modern artifacts over the past decades. In each of 57 cases examined, the particular combinations of demand elasticity and technical performance rate improvement are not consistent with dematerialization. Overall, the theory extension and empirical examination indicate that there is no dematerialization occurring even for cases of information technology with rapid technical progress. Thus, a fully passive policy stance that relies on unfettered technological change is not supported by our results.

## 1. Introduction:

Attempting to answer the basic underlying question and concern of sustainability – whether humans are taking more from the earth than the earth can safely yield- is the main objective underlying the concept of *dematerialization*. Malenbaum (1978) was one of the first researchers in this area and his key results are still among the most important. He utilized the concept of intensity of use defined as the ratio of the amount of materials (or energy) measured in bulk mass divided by GDP. When plotting intensity of use over time, he found "inverted U curves" peaking at different times in different countries (and for different materials) but at roughly a given GDP per capita for given materials. Also importantly, the peak intensity for a given material reached by subsequently developing countries decreases over time (relative to earlier developing countries). These two regularities are the essence of the conceptual basis for the "theory of dematerialization" according to Bernardini and Galli (1993). These authors speculate that the decreasing maximum intensity over time with usage of materials/energy per GDP might be a positive signal of a real dematerializing trend, but they eventually conclude that the empirical information at that time (1993) were insufficient to draw such a conclusion and suggest further examination of data.

Given the potential importance of the overall sustainability question, it is not surprising that there has been significant valuable work from the dematerialization perspective (see the next paragraph) and other perspectives, as for instance those claiming the urgent necessity of abating economic growth [the so-called 'degrowth' perspective, see for instance Knight et al (2013),, Turner (2008), Davidson et al (2014), and Lamb and Rao (2015)].,

From the dematerialization perspective, there has been significant work since Malenbaum. Dematerialization, is often defined as the reduction of the quantity of stuff and or energy needed to produce something useful and is then often assessed by a measure of intensity of use (consumption/production of energy and/or goods per GDP). Some of this research, Ausubel and Sladovick (1990) and Ausubel and Waggoner (2008), is encouraging emphasizing continuing decreases in consumption as a fraction of GDP. However, other researchers [Ayres (1995), Schaffartzik et al (2014), Senbel et al (2003), Allwood et al (2011), Gutowski et al (2013), Schandel and West (2010), Pulselli et al (2015)] are not as encouraging about continuation of economic growth with global dematerialization. Gutowski et al. (2013) call for much more attention to reducing the amount of material needed to fulfill a given function (referred to as "materials efficiency") and point out that decreasing usage of materials as a fraction of GDP is not sustainable unless absolute decreases in materials use occurs. The very recent and extensive work of Pulselli et al. (2015), presents a very interesting 3-dimensional analysis (resources, organization, and products/services) with which the authors scrutinize 99 national economies and conclude that no country is evidencing a dematerialization of economic activity, pointing out also that non-sustainable economic activity can take place over a wide range of income distributions.

There has also been extensive research on a closely related issue- usually called the Environmental Kuznets Curve (EKC). The EKC states that *emission of pollutants follow a inverted U curve as affluence increases*[1] and the concept was very positively viewed by some starting in the early 1990s [Grossman and Kreuger, (1991, 1994); IBRD, (1992)] as offering the strong possibility that emissions and pollution would not choke off economic growth but that economic growth might instead help eliminate pollution. However, the generality of the EKC has been seriously challenged on empirical, methodological and theoretical grounds [Stern et al (1996), Stern (2004), Kander (2005)].

---

[1] Although Kuznets did not discuss pollution or emission effects, his name is used since he postulated a similar inverted U shape for income-inequality as a function of affluence- GDP per capita.

Although the two issues, dematerialization and EKC analysis, differ in what is being considered, many fundamental issues are similar if not equivalent. Both discuss inverted U curves (in the first case of materials usage per capita, and in the latter of emissions) as affluence (GDP per capita) increases. Indeed, the term EKC has been also applied to dematerialization research (Canas et al, 2003) and a fundamental linkage was discussed by Kander (2005):
> "However, it is in principle true that economic growth may be reconciled with environmental concerns if dematerialization takes place."

Kander also establishes a strong base for skepticism concerning a suggested cause of such inverted U curves. She shows that the transition to a service economy does not necessarily lead to less industrial production, and supports her argument theoretically (using Baumol's insight about service growth as a portion of the economy being due to smaller productivity gains than industrial production) and empirically using data from 1800 to 1980 for Sweden. Kander also suggests that the analysis of EKC by Stern (2004) depicting four potential "proximate variables" causing a potential EKC can be applied to the dematerialization issue:
1. Scale - generally if all else remains the same, increasing production leads to increased emissions/consumption;
2. Changes in output mix (for example to services or information technology from heavy industry can reduce materials consumption and emissions)
3. Changes in input mix (if $CO_2$ emissions are the concern, changing from coal to natural gas would have a major effect at any scale on emissions due to electrical energy use).
4. Technological change

From analysis of long term $CO_2$ and other data from Sweden, Kander concludes that item 2 is minimal (and in the wrong direction) and that the progress made in Sweden is at least partially due to item 3- politically determined changes in fuel mix.

The analysis in the present paper focuses on technological change (which Kander indicates may have also contributed to the Swedish EKC.). A key goal of the simple theoretical extension presented here is to allow a broad set of cases to be examined concerning the absolute level of dematerialization achieved.
The theory of dematerialization is extended by explicit consideration of the ongoing technical progress on dematerialization. We do not treat substitution among technologies in this simple extension nor structural change in the economy but only the direct effect of technological change over long periods of time. However to do this requires that we also consider another highly researched issue- rebound, more widely known as the Jevons paradox.

The Jevons' paradox was first studied by Jevons (1865) and asserts that *energy* use is increased rather than decreased when more efficient energy technologies are introduced. This "paradox" is also known as the Khazzoom-Brooks postulate [Khazzoom (1980), Brooks (1984, 2000)], is also sometimes called backfire, and sometimes take back as well as rebound. The terminology is complex partly since an important issue is how much of the energy efficiency is essentially overwhelmed by increased energy consumption (backfire is the term used when improved energy efficiency results in increased (rather than decreased) energy consumption. Jevons as well as Khazzoom, Brooks and others argue that this strong effect is inevitable. In this paper, we are essentially adding some new approaches to examining whether technological progress relative to material usage does or does not lead to backfire for materialization- that is whether improvement in technical performance over time increases rather than decreases material consumption on an absolute basis. Davidson et al (2014) identify this issue in their analysis of the increasing impact of resource use over time (which they refer to as the 'effort factor'). Although there have been and continue to be authors who deny the rebound effect (especially the strongest or backfire result), there has been extensive theoretical work showing that the effect (Khazzoom-Brookes or Jevons) is at least a reasonable hypothesis (Saunders, 2000, 2005, 2008) and various systemic studies [Alcott, (2005), Sorrell, (2009), Schaffartzik et al (2014),] have tended to support the reality of such effects. However, section 4 in Sorrell (2009) opens with the following statement:

> "Time-series data such as that presented in Table 1[2] are difficult to obtain, which partly explains why relatively little research has investigated the causal links."

The second part of the work reported here is expands the number of empirical cases (time series data) that can be analyzed. Although the additional cases involve materials and technologies, they may have wider interest concerning the interplay of technological progress and rebound. Since energy is arguably more important to the economy than specific diverse materials (Sorrell, 2009), dematerialization in specific materials should be possible even if backfire occurs generally for energy technology. On the other hand, if rebound overcomes technological progress in numerous specific dematerialization cases, Jevons' paradox and authors who have supported it receive additional supporting evidence.

## 2. Dematerialization theory extension

As stated before, in this work we extend the theory of dematerialization by explicit consideration of two important factors that can enhance and/or mitigate the dematerialization process: i – the ongoing improvement in technical performance; ii – the rebound effect. We only consider cases of specific materials (or physical devices) and whether technological progress leads to an actual decrease over time in utilization of the materials.

---

[2] referring to lighting data from the UK given by Fouquet and Pearson (2006)

In order to analyze dematerialization quantitatively the following measures will be considered:

1 – the rate of change of per capita materials consumption – $dm_c/dt$ or $dm_{ci}/dt$ for a specific material, where $c$ denotes the per capita measure and $i$ some specific material/technology.

2 – the rate of population growth – $dp/dt$

3 – the rate of growth of GDP per capita – $dG_c/dt$

4 – the yearly *relative* increase of technological performance, defined as $k$ and as $k_i$ for a specific technology, $i$.

5 – the demand income elasticity $\varepsilon_{di}$ for goods and services, defined as relative increase in consumption of $i$ divided by the relative increase in national income

6 – the demand price elasticity, $\varepsilon_{dpi}$ is the relative increase in consumption of $i$ divided by the relative decrease in price of the good or service

7 – the rate of change of cost of a good or service with time, $dc_i/dt$, the rate of change of the performance of the good or service with time, $dq_i/dt$ and the rate of change of demand for a good or service with time, $dD_i/dt$.

**2.1 Incorporation of technological progress**

Technical progress is represented in this paper by the change in performance of technical artifacts as a function of time. Performance is measured by metrics that describe the effectiveness of a technology for a user/purchaser and have the same form as a generalized productivity measure (output/constraint) [Koh and Magee (2006), Magee et al (2014)]. By considering changes in performance over time rather than one-time improvements, this model is a more realistic treatment of technical change than some more sophisticated economic theories (for example, Saunders, 2008) that consider various production functions but consider technical change as a one-time delta. Thus, the model we propose is quite simple from an economics perspective but is arguably more advanced from the viewpoint of incorporation of technological progress.

Our treatment of technical performance change (technical progress) represents all such changes as occurring in metrics that either increase the performance or decrease the price of a technical artifact *exponentially* with time. This generalization of Moore's Law (Moore 1965) is

$$\frac{q_i}{c_i} = \frac{q_{i0}}{c_{i0}} \exp(k_i . t) \qquad (1)$$

where $q_i$ is the performance associated with use of $i$, $C_i$ is cost of $i$, $q_{i0}$ and $C_{i0}$ are the performance and cost at $t = 0$, and $k_i$ the relative annual increase in a specific ($i$) technical performance.

Thus, the relative performance (relative cost) of a given good or service $i$ increases (decreases) exponentially with time. There is extensive empirical evidence for such generalizations of Moore's Law being widely followed [Moore (2006), Martino (1971), Nordhaus (1997), Koh and Magee (2006), Nordhaus (2007), Koh and Magee, (2008), Koomey et al (2011)]. Two recent papers are also particularly noteworthy. Magee et al (2014) have looked generally at methodological issues involved with quantitative empirical study of technical performance trends and found that Moore's Law generally holds for performance over time whether performance does or does not include cost. They also found Moore's law to be a statistically satisfactory description for 71 different metric choices in 28 technological domains and more fundamentally appropriate for describing technical progress than other formulations based upon effort in a domain. Most importantly for this paper, Nagy et al (2013) have statistically examined 62 cases (considering only cost while holding performance constant) and found general support for equation (1). Since the cases considered in this paper all come from this reference, our use of this generalized Moore's law is an appropriate choice for quantifying technical progress. Relating this robust description of technical change to dematerialization is now required.

Since performance ($q_i$) is assessed by metrics that describe the effectiveness of a technology for a user/purchaser, the metrics have the form output/constraint with the constraint directly related to the amount of material used; performance is inversely proportional to materials used. For examples that follow equation 1, one can see from the metrics following the equal sign that the *materials* used: 1) to store a given amount of information [metric = mbits/cm$^3$ –see Koh and Magee, (2006)], 2) to perform a given amount of computation (MIPS/cm$^3$) –see Koh and Magee (2006) or 3) to store a given amount of energy (watt-hours/kg)- see Koh and Magee (2008), *all decrease* as the metric improves (or as technical performance increases). In fact, with such metrics, equation 1 shows the usage of materials to fulfill a given function decreasing as the technology improves exponentially by a constant ratio $k_i$ per year.

In other words, technical performance change described by equation (1) results in a given function being delivered with less material specifically as

$$d \ln m_{ci} /dt = - k_i \qquad (2)$$

where $m_{ci}$ is per capita usage of material $i$ and $k_i$ the annual rate of change of the relevant performance [3]. Equation 2 quantifies the point that more effective technologies result in reduced materials requirements. Allwood et al (2011) and Gutowski et al (2013) introduced the important concept of "materials efficiency" which measures the amount of material to *achieve a given level of function* (they use the term service) in a downstream artifact or service. Equation 2 gives a quantitative formulation of that concept. While this result seems to support many technological optimists (Diamandis and Kotker, 2012; Kaku, 2011; Brynjolfson and McAfee, 2014; Chertow, 2000; Waggoner and Ausubel 2008) who are well aware of the generality of Moore's Law, consideration of the rebound effect in section 2.2 will act to reverse this apparent support. Before introducing the rebound effect, we consider the influence of population on dematerialization

The key to analysis of dematerialization in specific cases is the measure $dm_i/dt$ which is the time rate of change of total usage (in mass or volume) of a specific material class *i*. The condition for *absolute* (this is appropriate because sustainability is an extensive not intensive issue as noted by Pulselli et al -2015) dematerialization in regard to *i* is that the usage of the material ($m_i$) must decrease with time[4]. Since materials use is simply population times the per capita materials usage, ($p$) x $m_{ci}$, one obtains decreasing $m_i$ over time if the relative rate of population growth is exceeded by the relative (decreasing) rate of per capita usage of a given material, or

$$\frac{1}{p} \times \frac{dp}{dt} + \frac{1}{m_{ci}} \times \frac{dm_{ci}}{dt} < 0 \qquad (3)$$

Stating this equation in log form, the criterion for dematerialization is then:
$$\left|\frac{d\ln m_{ci}}{dt}\right| > \left|\frac{d \ln p}{dt}\right| \qquad (4)$$

Considering that the world population is still increasing, even if at a lower rate, the *strong dematerialization criterion* means that the decrease in per capita use due to technical progress and given in equation (2) must exceed the positive increase of population growth.

---

[3] To illustrate specifically for the function (or service) of information storage, $m_{ci}$ is the per capita material used to store information and $k_i$ is the annual rate of increase in information storage technical performance.

[4] Davidson et al (2014) point out that lower quality ores may result in growth of environmental harm over time even with constant materials use and point to a possible technological improvement factor that might obviate this effect- none of this is considered here.

## 2.2 Incorporation of rebound

Equation 2 gives an estimate of the "materials efficiency" change with time without considering rebound. However, in the same time period, the rebound effect due to increases in q/c (purchasers opt for more function as the effective price decreases) offsets material usage decrease by $k_i$ x $\varepsilon_{dpi}$ which represents material that must be added back as technology improves simply because the technology then has more value to the user/purchaser and is therefore more highly used. In addition, the amount of material used increases due to economic growth (through increased consumption of function) which is given by $\varepsilon_{di}$ x $dlnG_c/dt$. Thus, considering rebound and economic growth, equation 2 becomes:

$$\frac{dln\, m_{ci}}{dt} = -k_i + \varepsilon_{dpi}k_i + \varepsilon_{di}\frac{dlnG_c}{dt} \qquad (5)$$

Equation 5 gives the change in a specific materials per capita consumption taking into account the combined effect of the yearly increase of technological performance ($k_i$), the rebound effect ($\varepsilon_{dpi}$ x $k_i$), and the effect of economic growth $\varepsilon_{di} \times \frac{dln\, G_c}{dt}$. Given constant output ($dG_c/dt = 0$), the annual relative change in per capita materials usage simply equals minus the relative change in annual technical performance ($k_i$) plus the rebound effect ($\varepsilon_{dpi} k_i$).

In order to allow for the possibility of economic growth, we make our second simplifying assumption that the demand elasticity for price and income are equal[5] and substituting equation (5) into inequality (4) we get for absolute dematerialization that:

$$\frac{d\, ln\, p}{dt} - k_i + \varepsilon_{di} \times k_i + \varepsilon_{di} \times \frac{dln\, G_c}{dt} < 0$$

or

$$\frac{d\, lnp}{dt} - k_i + \varepsilon_{di}\left(k_i + \frac{dln\, G_c}{dt}\right) < 0 \qquad (6)$$

---

[5] This is only roughly justified by assumption that relative increases in usage due to increased value (decreased price or increased function) are the same as the relative increase in usage due to increases in income. It allows us to leave the potential for economic growth in the model so it is a useful assumption that might be removed in a less simple model.

Inequality (6) contains specific relationships for all items in the IPAT identity[6] [Ehrlich and Holden (1970), Commoner et al, (1971)]: I is impact (or materials usage in the case of dematerialization) which is growing if the left hand side of the inequality is positive, P is population and $\frac{dlnp}{dt}$ is the time dependence or growth rate of population, A is affluence and $\frac{dlnG_c}{dt}$ is the time dependence of affluence, T is technology and $k_i$ is the time dependence of technological performance. Inequality 6 can thus be termed as "in the IPAT framework" but is explicit about relationships over time among the terms and includes rebound which is not explicitly in the IPAT framework. Moreover, our approach differs from more recent derivatives of IPAT such as STIRPAT (York et al, 2003, and Liddle, 2015) which although testable (IPAT is not) treat technology (T) as a residual. If we were going to use an acronym for our model showing links to IPAT, we might suggest IPAT$\varepsilon$k.

## 3. Graphical representation

In inequality (6), $\frac{dlnp}{dt}$ and $\frac{dln\, G_c}{dt}$ are variables that can be obtained from available time series data on the growth of population and the growth of GDP. $k_i$ is a complex measure that is different for different families of technologies (but constant over time for each case) and will be given for cases later in this paper; $k_i$ has been found to be in the range of 3-65% per year (Magee et al, 2014) for different technological domains. Finally, $\varepsilon_{di}$ is complex but can be estimated for specific cases and will also be considered in the cases covered later in this paper. Before undertaking empirical examination, it is useful to show graphically how the fundamental parameters ($k_i$ and $\varepsilon_{di}$) delineate what is possible relative to dematerialization.

Figure 1 below depicts the time dependence (last 50 years) of the two "less-complex" terms of inequality (6), namely $\frac{dlnp}{dt} + \varepsilon_{di} \times \frac{dln\, G_c}{dt}$ assuming $\varepsilon_{di}$ = 0.5, which represents an approximate value for artifacts that are evidencing declining rates of demand as a ratio of GDP. Figure 1 demonstrates that the sum of the non-rebound growth terms exhibits a declining linear trend that favors dematerialization emerging over time.

---

[6] This is also referred to as the Commoner-Ehrlich equation.

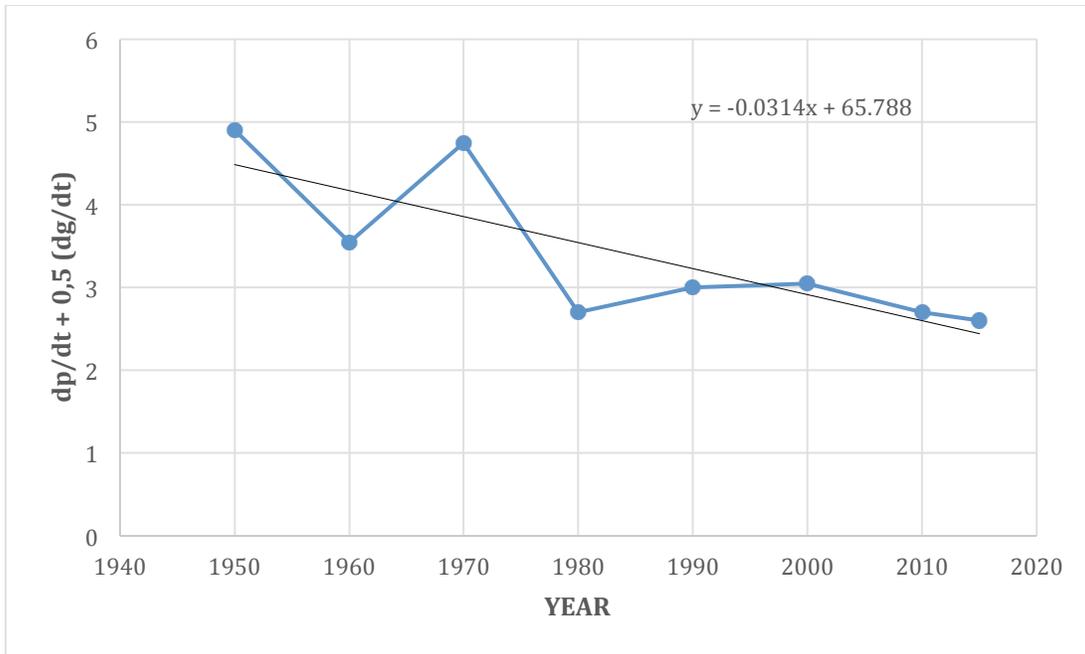

Figure 1: Trends over time in population growth + 0.5x GDP growth

We now turn to examining the effect of key variables on dematerialization by showing the boundary defined by inequality 6 as a function of the variables. The next three graphs show the areas of materialization and dematerialization for some possible values of $\varepsilon_{di}$ and $k_i$, and for approximate current values of $\frac{d\ln p}{dt}$ and $\frac{d\ln G_c}{dt}$ (0.01 and 0.03 respectively). Figure 2 shows that dematerialization occurs (under the somewhat reasonable assumption of $k_i$ = 0.05 and $\varepsilon_{di}$ = 0.5) in the lower left triangle bounded by a maximum GDP growth of 5% per year and a max population growth of 2.5%. This result is somewhat encouraging by indicating the possibility of achieving economic growth while dematerializing.

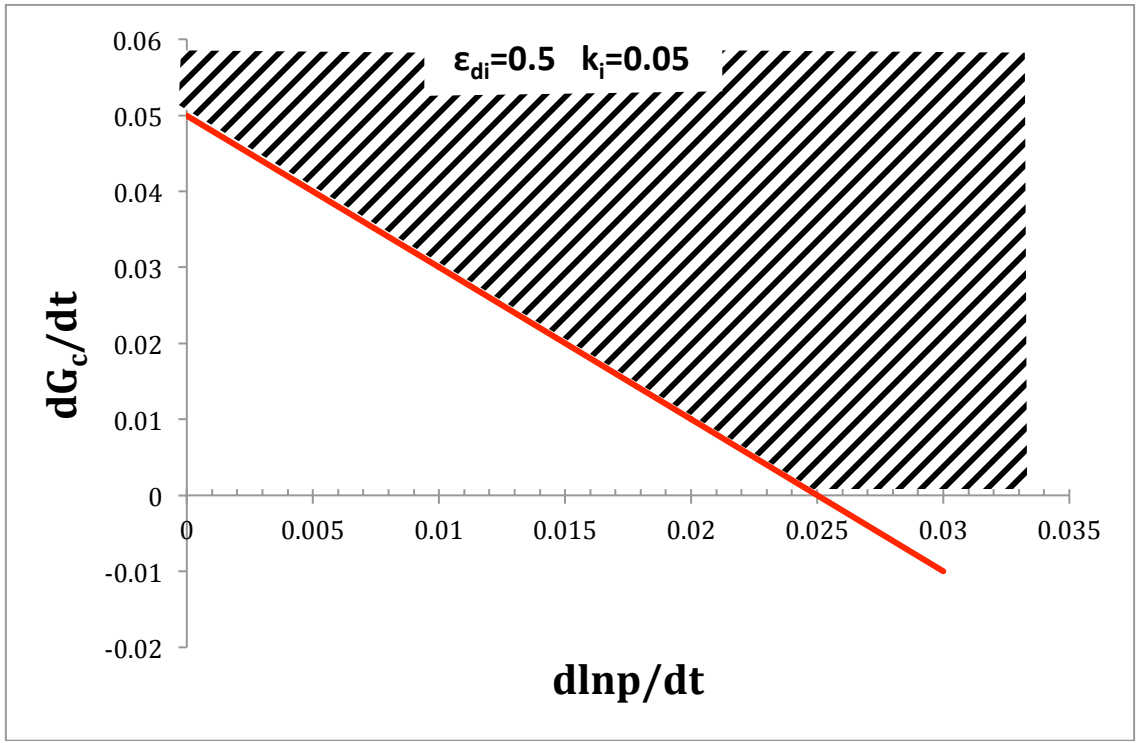

-Dematerialization    -Materialization

Figure 2: materialization and dematerialization for fixed demand elasticity and population growth for various values of GDP growth and population growth

Figure 3 is even more encouraging as it shows a large dematerialization region at high (but not unreasonable) $k_i$ values when $\varepsilon_{di}$ = 0.5 and population growth is 1% per year. In this instance, *much higher economic growth with dematerialization* is possible (10% or more) at $k_i$ = 0.15 and beyond showing apparently substantial growth potential with higher rates of technical improvement. However, the encouragement offered by Figures 2 and 3 is strongly countered by the fact that $\varepsilon_{di}$ is perhaps even more important than $k_i$. This is shown by Figure 4 where all possible values of $k_i$ and $\varepsilon_{di}$ are shown assuming actual values for population and economic growth. For all values of $\varepsilon_{di}$ greater than or equal to 1, *no dematerialization* is possible for any value of $k_i$ . These results suggest that Engel's Law[7] must operate for dematerialization since it only holds when $\varepsilon_{di}$ is less than 1.

---

[7] Engel's law is that agricultural product share of GDP decreases for all societies moving beyond subsistence. It is often generalized to indicate that all commodities have demand elasticity less than 1.0.

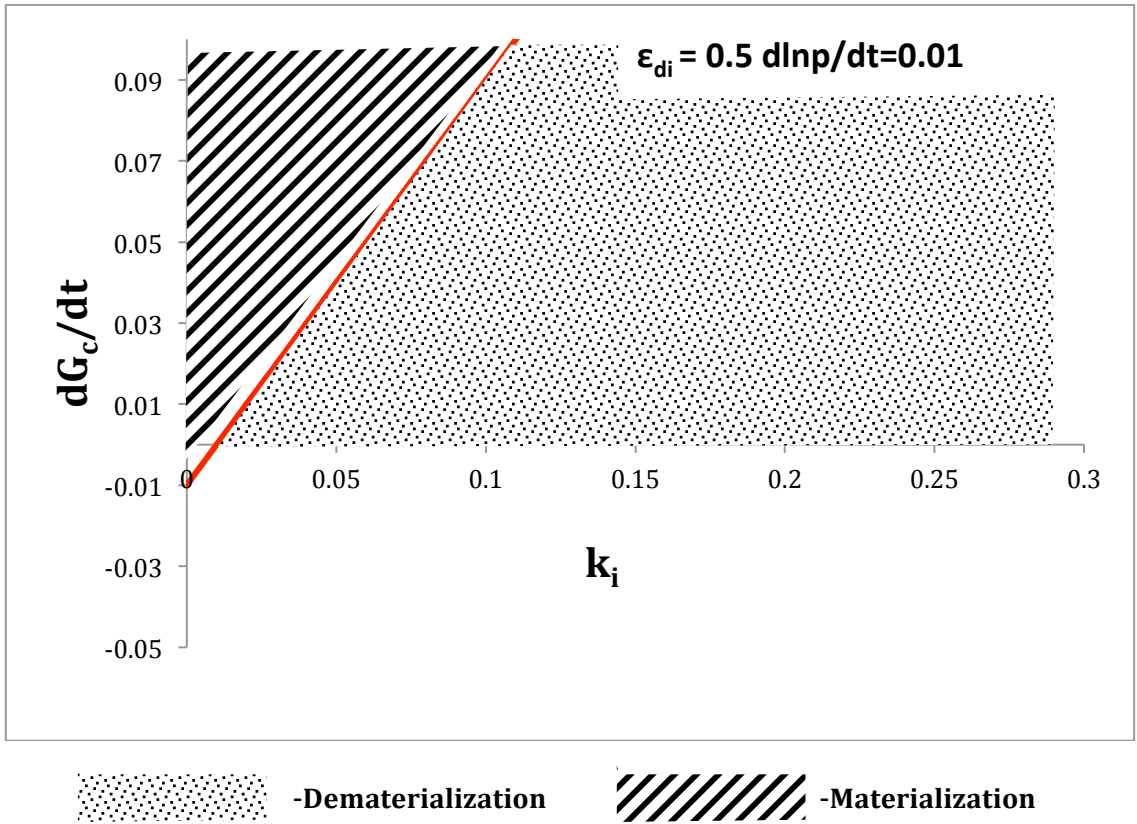

-Dematerialization    -Materialization

Figure 3: Materialization and dematerialization for various levels of economic growth and technical capability improvement rate at population growth of 1% per year and demand elasticity = 0.5.

Our extension of dematerialization theory to include technical performance and the rebound effect shows the extreme importance of $k_i$ and $\varepsilon_{di}$ in assessing the feasibility of dematerialization with economic growth. The importance of demand elasticity offsetting performance improvement is implicit in Jevons, Khazzoum, Brookes and others. Complementing this past work, the simple graphical representation (Figures 3 and 4) adds to understanding how the key processes of technological improvement and the rebound effect exert large influence on the potential for dematerialization with economic growth. In doing so, the model also specifies the assumptions to arrive at the results. We do not presume that answers to the key questions are thereby known- empirical results are still necessary even to assess the specific predictions of this simple model. A major challenge is to prescribe values for $k_i$ and $\varepsilon_{di}$. The next section of the paper develops a new approach for estimating $\varepsilon_{di}$: this method and a key recent data-rich paper [Nagy et al (2013)] allows estimates for $k_i$ and $\varepsilon_{di}$ to be made for a large number of cases. The key empirical contribution of this paper is to examine the most relevant 57 of these 62 cases in light of the dematerialization criteria given in inequality 6 (which defines the dematerialization region in Figure 4). This involves mapping all of the 57 cases onto plots such as Figure 4 in order to determine if they are either in the materialization region or the dematerialization region.

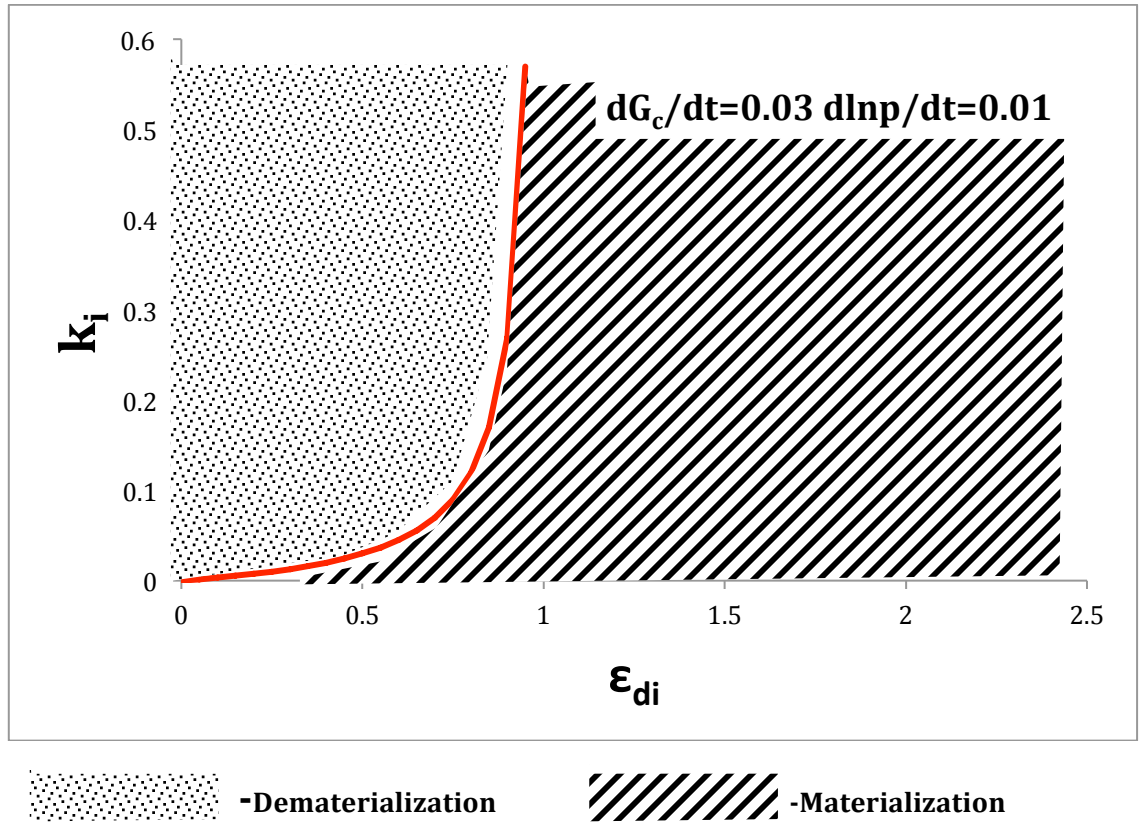

Figure 4 Materialization and dematerialization at values of $k_i$ and $\varepsilon_{di.}$

## 4. $k_i$ and $\varepsilon_{di}$ estimation method

Nagy et al (2013) have examined 62 cases of changes in prices and production/demand as a function of time. For all cases, Nagy et al found exponential relationships between price and time as well as production/demand with time. The authors report the exponent in these relationships in their Supplemental Information. The key relationships are:

$$c_i = c_0 \exp(-k_i t)$$
$$D_i = D_{0i} \exp(g_i t) \qquad (8)$$

Since price/cost ($c$) at constant function is an inverted metric for technological improvement, fits to the first equation directly yield an estimate of $k_i$[8]. More importantly, the exponent for the demand exponential ($g_i$) can be used to estimate $\varepsilon_{di}$ for each of the 62 cases as will now be shown. We can write $g_i$ as the total (logarithmic) derivative of demand with respect to time and examine its decomposition into dependence on $G_c$ (still GDP per capita) and $c_i$ (price) since $G_c$ and $c_i$ are both separately dependent upon time. We have:

$$g_i = \frac{d \ln D_i}{dt} = \frac{\partial \ln D_i}{\partial \ln G_c} \cdot \frac{\partial \ln G_c}{\partial t} + \frac{\partial \ln D_i}{\partial \ln c_i} \cdot \frac{\partial \ln c_i}{\partial t}$$

(9)

---

[8] Called m by Nagy et al in their paper; we also note that Nagy et al report g and m in their SI on a log 10 basis and these are converted in our Tables 1 and 2 to natural logs consistent with Equation 8 (and their equation 9 as well).

The right hand side of this equation has two terms both of which are products of two partial derivatives. The first term is the income elasticity of demand, $\varepsilon_{di}$ times the growth rate of $G_c$ and the second term[9] is the price elasticity of demand $\varepsilon_{dpi}$ multiplied by $k_i$. If we again conveniently take the demand elasticities as equal (and constant over time), we have

$$g_i = \varepsilon_{di}\left(\frac{d \ln G_c}{dt} + k_i\right)$$

(10)

This can be rearranged to find $\varepsilon_{di}$ from known quantities (using $g_i$ and $k_i$ from Nagy et al and $\ln G_c$ as a function of time, $t$, from the World Bank, 2012) as

$$\varepsilon_{di} = \frac{g_i}{\left(k_i + \frac{d \ln G_c}{dt}\right)}$$

(11)

## 5. Results

### 5.1 Key variables and mapping onto formalism

The estimates of $\varepsilon_{di}$ (and the range of years for the data and the values of $k_i$ and $g_i$ from Nagy et al, 2013) are given in Table 1 and 2 for the 57 cases (of the 62 in Nagy et al) most relevant to the issues in this paper. Table 1 is for the chemicals category as labeled by Nagy et al and Table 2 includes the hardware and energy industry cases. For this paper, it is useful to note that Table 1 is most relevant for dematerialization and that the energy technologies in Table 2 add cases for consideration of energy –directly relevant to the Jevon's paradox. The hardware cases in Table 2 represent more rapidly improving modern technological products.

---

[9] Two negative signs in the second term are not shown as their product is positive.

**Table 1 For Chemical technologies: Values of $g_i$ and $k_i$ from Nagy et al (2013), values of $\varepsilon_{di}$ calculated from Eq. 11 and the dematerialization value from inequality 6.**

| Technology Chemicals | Time period | $g_i$ | $k_i$ | $\varepsilon_{di}$ | Inequality 6 |
|---|---|---|---|---|---|
| AcrylicFiber | 1960-1972 | 0,176744 | 0,104651 | 1,142857 | 0,092093 |
| Acrylonitrile | 1959-1972 | 0,17907 | 0,076744 | 1,412844 | 0,122326 |
| Aluminum | 1956-1972 | 0,081395 | 0,009302 | 1,372549 | 0,092093 |
| Ammonia | 1960-1972 | 0,109302 | 0,090698 | 0,77686 | 0,038605 |
| Aniline | 1961-1972 | 0,062791 | 0,05814 | 0,580645 | 0,024651 |
| Benzene | 1953-1968 | 0,083721 | 0,062791 | 0,742268 | 0,04093 |
| BisphenolA | 1959-1972 | 0,151163 | 0,062791 | 1,340206 | 0,108372 |
| Caprolactam | 1962-1972 | 0,213953 | 0,116279 | 1,286713 | 0,117674 |
| CarbonDisulfide | 1963-1972 | 0,044186 | 0,02093 | 0,622951 | 0,043256 |
| Cyclohexane | 1956-1972 | 0,139535 | 0,053488 | 1,348315 | 0,106047 |
| Ethanolamine | 1955-1972 | 0,113953 | 0,062791 | 1,010309 | 0,071163 |
| EthylAlcohol | 1958-1972 | 0,072093 | 0,013953 | 1,127273 | 0,07814 |
| Ethylene | 1954-1968 | 0,193023 | 0,037209 | 2,213333 | 0,175814 |
| Ethylene2 | 1960-1972 | 0,134884 | 0,065116 | 1,171717 | 0,089767 |
| EthyleneGlycol | 1960-1972 | 0,095349 | 0,067442 | 0,811881 | 0,047907 |
| Formaldehyde | 1962-1972 | 0,095349 | 0,060465 | 0,863158 | 0,054884 |
| HydrofluoricAcid | 1962-1972 | 0,081395 | 0,002326 | 1,555556 | 0,09907 |
| LDPolyethylene | 1953-1968 | 0,255814 | 0,102326 | 1,679389 | 0,173488 |
| Magnesium | 1954-1972 | 0,051163 | 0,006977 | 0,897959 | 0,064186 |
| MaleicAnhydride | 1959-1972 | 0,127907 | 0,055814 | 1,208791 | 0,092093 |
| Methanol | 1957-1972 | 0,088372 | 0,05814 | 0,817204 | 0,050233 |
| NeopreneRubber | 1960-1972 | 0,076744 | 0,02093 | 1,081967 | 0,075814 |
| Paraxylene | 1958-1968 | 0,232558 | 0,1 | 1,550388 | 0,152558 |
| Pentaerythritol | 1952-1972 | 0,090698 | 0,04186 | 0,987342 | 0,068837 |
| Phenol | 1959-1972 | 0,097674 | 0,081395 | 0,743363 | 0,036279 |
| PhtalicAnhydride | 1955-1972 | 0,081395 | 0,072093 | 0,666667 | 0,029302 |
| PolyesterFiber | 1960-1972 | 0,27907 | 0,137209 | 1,490683 | 0,16186 |
| PolyethyleneHD | 1958-1972 | 0,216279 | 0,097674 | 1,464567 | 0,138605 |
| PolyethyleneLD | 1958-1972 | 0,17907 | 0,088372 | 1,294118 | 0,110698 |
| Polystyrene | 1944-1968 | 0,2 | 0,05814 | 1,849462 | 0,16186 |
| Polyvinilchloride | 1947-1968 | 0,169767 | 0,076744 | 1,33945 | 0,113023 |
| PrimaryAluminum | 1930-1968 | 0,102326 | 0,025581 | 1,353846 | 0,096744 |
| PrimaryMagnesium | 1930-1968 | 0,174419 | 0,025581 | 2,307692 | 0,168837 |
| Sodium | 1957-1972 | 0,032558 | 0,016279 | 0,491228 | 0,036279 |
| SodiumChlorate | 1958-1972 | 0,1 | 0,039535 | 1,116883 | 0,080465 |
| Styrene | 1958-1972 | 0,118605 | 0,069767 | 0,990291 | 0,068837 |
| TitaniumSponge | 1951-1968 | 0,27907 | 0,116279 | 1,678322 | 0,182791 |

| Urea | 1961-1972 | 0,151163 | 0,074419 | 1,214953 | 0,096744 |
| VinylAcetate | 1960-1972 | 0,127907 | 0,076744 | 1,009174 | 0,071163 |
| VinylChloride | 1962-1972 | 0,14186 | 0,090698 | 1,008264 | 0,071163 |

**Table 2: For Hardware and Energy technologies: Values of $g_i$ and $k_i$ from Nagy et al (2013), values of $\varepsilon_{di}$ calculated from Eq. 11 and the dematerialization value from inequality 6.**

| Technology Hardware Ind. | Time period | $g_i$ | $k_i$ | $\varepsilon_{di}$ | Inequality 6 |
|---|---|---|---|---|---|
| DRAM | 1972-2007 | 0,604651 | 0,44186 | 1,281419 | 0,182791 |
| HardDiskDrive | 1989-2007 | 0,651163 | 0,651163 | 0,955958 | 0,02 |
| LaserDiode | 1983-1994 | 0,744186 | 0,325581 | 2,092871 | 0,438605 |
| Transistor | 1969-2005 | 0,488372 | 0,488372 | 0,942127 | 0,02 |
| **Technology Energy Ind.** | **Time period** | $g_i$ | $k_i$ | $\varepsilon_{di}$ | **Inequality 6** |
| CCGTElectricity | 1987-1996 | 0,174419 | 0,02093 | 3,424658 | 0,173488 |
| CrudeOil | 1947-1968 | 0,05814 | 0,009302 | 0,980392 | 0,068837 |
| ElectricPower | 1940-1968 | 0,106977 | 0,037209 | 1,226667 | 0,089767 |
| Ethanol | 1981-2004 | 0,139535 | 0,053488 | 1,671309 | 0,106047 |
| GeothermalElectr | 1980-2005 | 0,097674 | 0,051163 | 1,203438 | 0,066512 |
| MotorGasoline | 1947-1968 | 0,065116 | 0,013953 | 1,018182 | 0,071163 |
| OffshoreGasPipel. | 1985-1995 | 0,255814 | 0,113953 | 1,77706 | 0,16186 |
| OnshoreGasPipel. | 1980-1992 | 0,15814 | 0,016279 | 3,417085 | 0,16186 |
| Photovoltaics1 | 1976-2003 | 0,225581 | 0,065116 | 2,371638 | 0,180465 |
| Photovoltaics2 | 1977-2009 | 0,213953 | 0,104651 | 1,588946 | 0,129302 |
| WindElectricity | 1984-2005 | 0,44186 | 0,093023 | 3,591682 | 0,368837 |
| WindTurbine1 | 1982-2000 | 0,27907 | 0,04186 | 3,883495 | 0,257209 |
| WindTurbine2 | 1988-2000 | 0,534884 | 0,039535 | 7,692308 | 0,515349 |

Figure 5 shows the 57 cases in Tables 1 and 2 mapped onto the format of Figure 4. The $k_i$ and $\varepsilon_{di}$ values for each of the individual lines in the Tables become a point in Figure 5a (chemicals), Figure 5b (hardware) or Figure 5c (energy). Since $dP/dt$ and $dG_c/dt$ are not precisely constant over time, the dematerialization boundary for figures 5a and 5c are drawn for approximate $dG_c/dt$ and $dP/dt$ for the 1940s through 1960s whereas figures 5b is consistent with Figure 4 and is applicable for the 1980s onward. Earlier dated cases are placed on Figure 5a (the chemical cases from Table 1) and figure 5c (energy cases from Table 2) where the dematerialization border is at higher values of $k_i$. The more recent hardware cases from Table 2 are mapped onto Figure 5b. Examining Figures 5a, 5b and 5c, none of the 57 cases are in the dematerializing region. The last column of Table 1 shows the actual value for inequality 6 for each individual chemicals case. Table 2 shows the actual values for the hardware and energy industry cases. None of the values are less than zero so *none* are reducing in material usage in the periods for which the times series data from Nagy et al apply and thus none are calculated as dematerializing consistent with Figure 5.

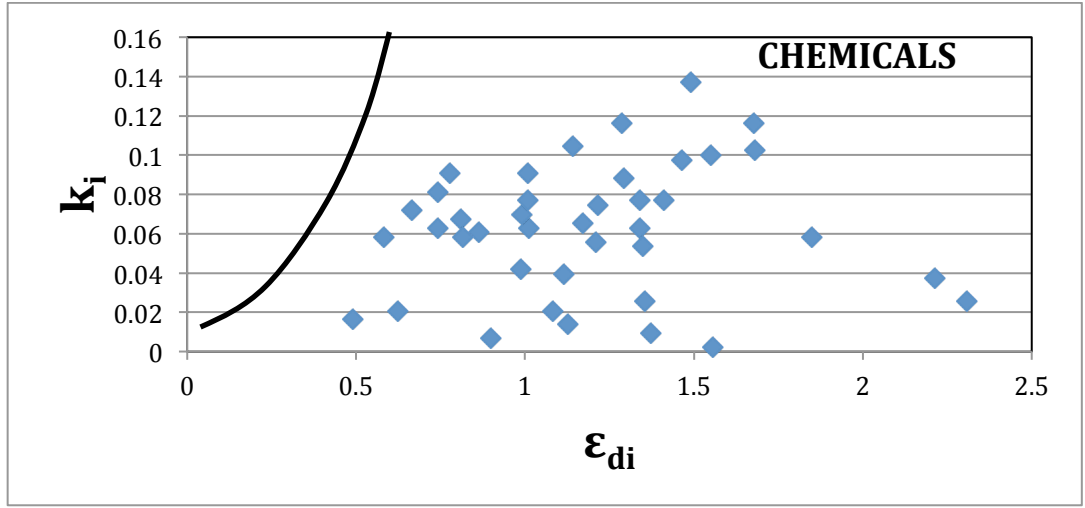

Figure 5a: All chemical technology cases from Table 1 plotted in the format of figure 4 but for values of population growth and GDP growth consistent with the time frame of the chemical technologies data.

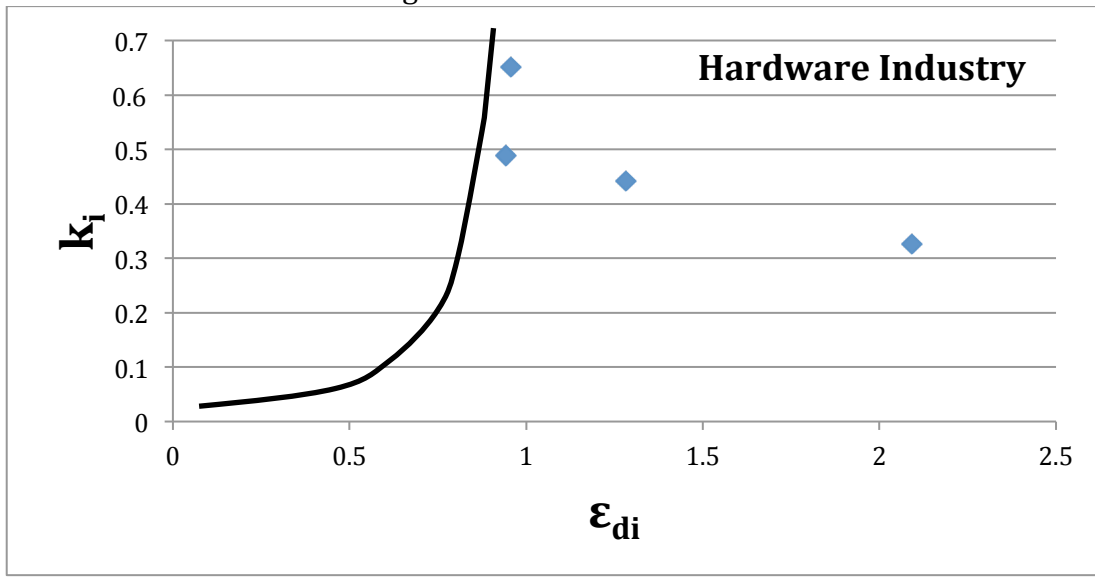

Figure 5b: The hardware technology cases from Table 2 plotted in the format of Figure 4

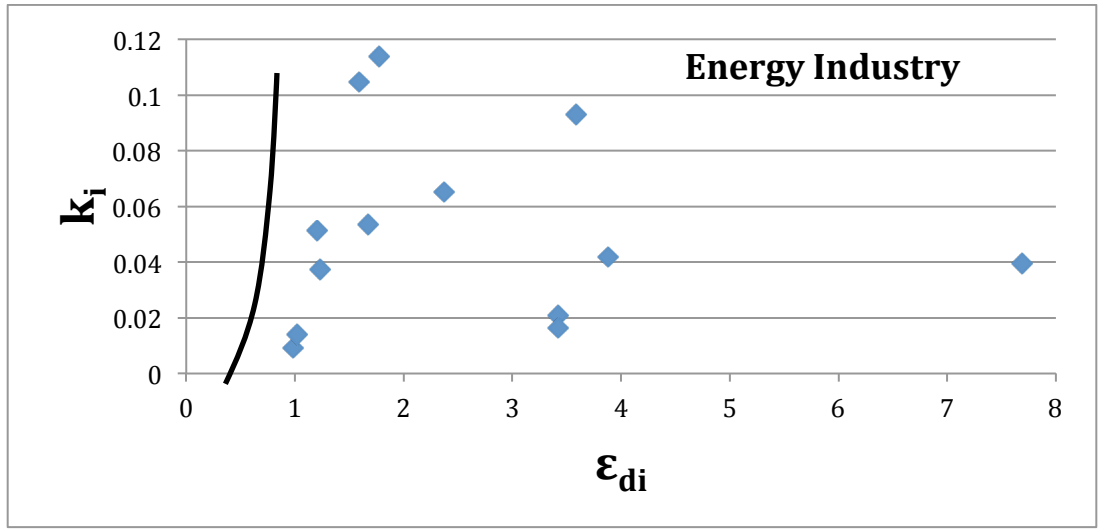

Figure 5c: The energy technology cases from Table 2 plotted in the format of Figure 4 with values for population growth and economic growth consistent with the time frame for the energy technology data.

Absolute dematerialization requires high enough $k_i$ and low $\varepsilon_{di}$. The modern product technologies shown in Figure 5b all have significantly higher $k_i$ values (> 0.3) but nonetheless are in the materialization range since all also exhibit relatively high $\varepsilon_{di}$ values (> 0.9). Several chemical (materials) technologies have low demand elasticity; however, the lowest $\varepsilon_{di}$ materials (Aniline, CarbDisulf, Sodium) have very low $k_i$. Thus these cases also fall into the materialization region. These empirical results based upon a variety of time series suggest that absolute dematerialization is not easily achieved since the diversity and multiplicity of the 57 cases yields none that do.

**5. 2 Additional Results**
Although the breadth of cases from the Nagy et al data is impressive, the 40 materials cases (or chemicals using their terminology) all have time series whose latest dates are more than 4 decades ago. Thus, one concern is whether these results are good evidence of what may be occurring today. To explore this issue, we examined 69 materials cases from 1960 to 2010 using a variety of sources [Fibers: USDA , World Bank, CIRFS; Metals and Minerals: USGS; Cellulose/Wood – FAO; Plastics: PEMRG , H. G. Ellias (2003), and Kunststoff GmbH] These results show that 6 out of these 69 cases show an absolute decline in materials usage over the 50 year period potentially suggesting that some materials are now entering technologically-enabled absolute dematerialization. However, examining the six cases instead suggests that this is probably not the case. The 6 cases are: asbestos, beryllium, mercury, tellurium, thallium, and wool. Four of these are clearly not examples of technological improvement overcoming rebound leading to dematerialization but instead the dematerialization for asbestos, beryllium, mercury and thallium has occurred because of legal restrictions on their use due to toxicity issues. The other two cases – Tellurium and wool -are probably examples of substitution which is a major outstanding issue relative to dematerialization [Ruth (1998)], and it will be discussed further below.

## 6. Discussion

Although the breadth and number of cases is good evidence of the difficulty of achieving dematerialization for a broad range of technical performance improvement rates, there are limitations that suggest care in making too broad a generalization based upon our results. First, our economic model is simple essentially using demand elasticity as the mechanism for quantifying rebound. More in depth -but necessarily less broad analysis- is given in Liddle (2015) who gives robust estimates of elasticity of Carbon emissions with respect to population and income. Interesting future work would be to extend Liddle's analysis to include dematerializations cases.  Secondly, the method we developed for extracting elasticity from the time series data rely upon the assumption that demand elasticity due to income increases and the demand elasticity due to more attractive products are equal *and* constant over time. Balancing the simplicity of the economic model is the fact that we use (to our knowledge for the first time) a richer quantification of technical progress that is firmly based upon other empirical work (the generalized Moore's Law). Overall, it is our contention that this simple model is useful for three reasons: 1) because it leads to simple visualization (the graphical representation); 2) because the assumptions underlying the model are clear and 3) because it enabled broader empirical tests. Further modeling and empirical work should be able to probe the importance of the assumptions and the adequacy of the time series data we have used.

Despite the caveats just mentioned, the results shown in Figure 5 consider both technological change and the rebound effect and clearly show a challenge in relying on "automatic dematerialization" for the future that is consistent with empirical studies such as Schandel and West (2010). The results also indicate that "materials efficiency" through new designs and technology is not sufficient to obtain dematerialization. The significant increase in "materials efficiency" (reductions of needed material to achieve a given level of function) in the DRAM example will not often be surpassed, but this example (and the other few rapid improving material efficiency cases) still result in absolute materialization due to relatively high $\varepsilon_{di}$. When demand elasticity is near (or worse greater than) 1, dematerialization will not occur with any level of improvement in efficiency of materials usage. In regard to our desire to understand the combined effect of technical performance improvement and rebound, the results are at least highly suggestive. Results from previous multivariate correlation research [Steinberger et al (2010), York et al (2003)] correlating total industrial material consumption with $G_c$ indicate that income elasticity for overall material consumption is near to or greater than one. This "broad combination elasticity" is consistent with the results reported here for multiple disaggregated cases. Moreover, the analysis in Liddle (2015) improves on some earlier weaknesses in STIRPAT analyses and it also suggests high income elasticity for Carbon emissions. These results along with the analysis in this paper give further support to the overall low potential for dematerialization based upon unfettered technological progress. Continuation of work to find better $k_i$ and $\varepsilon_{di}$ values is certainly worth pursuing as is the development of more complex models. However, it seems likely to us that such work will support the major empirical finding reported here- that direct dematerialization due to technological progress will not occur. Further theory and empirical work might better focus on the remaining critical issues in dematerialization.

A major issue not addressed by our work is the issue of *substitution*. Our formulation of the rebound constraint (Jevons' paradox) does not consider substitution of materials, artifacts or functions and all are possible. Observing a decrease in material usage relative to GDP (or even an absolute decrease) for an old technology is of no help, if newer technologies substituting for it (or supplementing it) cause the total consumption to continue increasing. This would appear to be the case for wool (and probably tellurium) in its dematerialization. Synthetic fiber is one of the strongest growing material classes in the 69 we examined and the decrease in wool usage is more than counterbalanced by this growth. On the other hand, technological development does not only increase the performance of existing technologies but also results in the emergence of totally new technologies. If the new technologies use a very different resource base, technological development might be able to achieve success environmentally and economically [Ruth (1998), Kander (2005)]. However, it is also possible that the totally new technologies will be just as problematic as the outgoing technology. In the following paragraph, we qualitatively discuss a major case of sufficient breadth to introduce the full scope of the substitution issue relative to dematerialization.

The continuing rise of Si based semiconductors is perhaps the major technological fact of the past five or more decades. Silicon-based technology is a "general purpose technology" [Bresnahan and Trajtenberg (1995)] underlying much of the improvement in information storage, information transmission and computation since the 1960s and some have argued [Brynjolfsson and McAfee (2014)] that it is the most important general-purpose technology ever. From 1968 to 2005, the number of transistors sold for use has increased by $10^9$; by 2005 there were more transistors used then printed text characters (Moore, 2006)! However, the industry revenue per transistor has fallen almost as dramatically (Moore, 2006) as has the amount of material needed to make a transistor. Nonetheless, the usage of silicon has grown significantly since 1970. We find it has grown by 345% over this period but also find the growth is less than GDP growth (472% in the same period) and that much of the growth of Si usage is associated with non-electronic applications. This growth would be $10^5$ (or more) times as high if a 2005 transistor used as much Si as one manufactured in 1968 showing the importance of the profound change in "materials efficiency" for this technological domain[10].

For a general-purpose technology such as transistors, examination of substitution requires more than considering usage. Si-based technologies have enabled entire new industries such as wireless communication, the Internet, social networks, software systems and others. Each of these involves artifacts and systems that consume materials so the continuous rapid development of this technology has far broader implications on dematerialization than the use of Si. Moreover, a key question is to what extent these new technologies enabled by silicon have substituted for more energy and/or material intensive industries.

Two example questions are offered to clarify the complexity of the substitution issue. The first is to consider the potential for a changing basic function: substitution of electronic communication enabled "virtual" visits to replace travel. Although the communication technologies are not yet able to meet this desire (and it is not clear that it will ever be an adequate full substitute for "real" travel), if reversal in the rapid growth of long distance travel were to occur, it is likely (but would take careful study of the infrastructure and artifacts created and eliminated) that significant real dematerialization could occur. A second example is the growth of Si usage associated with Solar Photovoltaic: we find that this usage has now eclipsed electronic uses of Silicon. Since this application is essentially on a path to replace fossil fuel generation of electricity[11], [Devezas et al (2008)], the comparison would have to involve all the infrastructure and devices for both of these alternatives in order to determine the actual dematerialization. The significant reduction in $CO_2$ is – in this case- perhaps more important than the net materialization associated with the alternatives. Nonetheless, the consideration of the full impact of solar cells vs. fossil fuels on materialization would be quite complex on its own involving not only solar modules and fossil fuel generating plants but also needed electrical transmission and storage infrastructures, fossil fuel extraction systems, extraction systems for solar module materials, and many others to understand the materialization aspect of this one substitution being enabled at least partly by improvement in silicon-based technology.

## 7. Concluding Remarks

We believe that the theory/framework introduced in this paper clarifies the interaction of technological improvement with demand rebound in a simple but fairly useful manner. The framework and its application to 57 different cases clearly indicate that technological improvement has not resulted in "automatic" dematerialization in these cases. Moreover, the combination of high improvement rates with high demand elasticity seems to indicate that the future is not highly likely to reverse this finding. The results support the position of Jevons, Khazzoom and Brooks without recourse to a special role for energy in showing that rebound can (and apparently usually does) overcome technological progress as far as absolute dematerialization is concerned. The findings also provide support for the view (Stern, 2004) that environmental impact does not continue to diminish as affluence increases. An optimistic possibility yet remains: drastic substitution (on a functional and system basis) of more benign technologies where such technologies result from continuing technological change. The discussion of the silicon-enabled general-purpose technology here is qualitative and only a minimal outline. Nonetheless, this hopefully is sufficient to indicate the importance of theory and empirical efforts on substitution studies. A deeper understanding of substitution effects is also essential to enable effective policy design for dematerialization. With our current very limited knowledge about substitution, we have no reliable approach to developing policy relative to the effect of the major technology of the past 50 years. Reliable assessment is complex because semi-conductor technology [Kander (2005), Brynjolfsson and McAfee (2014)] has enabled so many other technologies that even an approximate global substitution study appears quite challenging.

---

[10] This counterfactual is somewhat misleading because the growth of usage would be much lower if the improvements had not occurred ("reverse rebound").

[11] We note that the promise for solar PV relative to fossil fuels is that the technical performance increase ($k$) is about 0.1 per year for solar PV [Benson and Magee (2014)] and less than 0.03 for fossil fuel energy systems [McNerney et al (2011)].

## 8. Acknowledgements

The first author (Magee) is grateful for support from the SUTD/MIT International Design Center, and the second author (Devezas) is grateful for support from the FCT through the Research Unit C.MAST.

PEMRG (Plastics Europe Market Research Group), http://www.europeanplasticsnews.com/marketdata/

Pulselli, F.M., Coscieme, L., Neri, L., Regoli, A., Sutton, P.C., Lemmi, A, and Bastianoni, S., 2015. The world economy in a cube: A more rational structural representation of sustainability. Global Environmental Change 35, 41 – 51.

Ruth, M.,1998. Dematerialization in five US metal sectors: implications for energy use and CO2 emissions. Resources Policy 24, 1- 18.

Saunders, H.D., 2000. Does predicted rebound depend on distinguishing between energy and energy services? Energy Policy 28 (6–7), 497–500.

Saunders, H.D., 2005. A calculator for energy consumption changes arising from new technologies. Topics in Economic Analysis and Policy 5 (1), 1–31.

Saunders, H.D., 2008. Fuel conserving (and using) production function. Energy Economics 30 (5), 2184–2235.

Schaffartzik, A., Mayer, A., Gingrich, S., Eisenmenger, N., Loy, C., and Krausmann, F., 2014. The global metabolic transition: Regional patterns and trends of global material flows, 1950 – 2010. Global Environmental Change 26, 87 – 97.

Schandl, H., and West, J., 2010. Resource use and resource efficiency in the Asia – Pacific region. Global Environmental Change 20, 636 – 647.

Senbel, M., McDaniels, T., and Dowlatabadi, H., 2003. The ecological footprint: A non-monetary metric of human consumption applied to North-America. Global Environmental Change 13, 83 – 100.

Sorrel, S., 2009. Jevons' paradx revisited: The evidence for backfire from improved energy efficiency. Energy Policy 37, 1456 – 1469.

Steinberger, J. K., Krausman, F., and Eisenmenger, N., 2010, Global patterns of materials use: A socioeconomic and geophysical analysis. Ecological Economics, 69, 1148-1158

Stern, D. I., Common, M. S., and Barbier, E. B., 1996. Economic growth and environmental degradation: The environmental Kuznets curve and sustainable development. World Development, 24, 1151–1160.

Stern, D. I., 2004, The rise and fall of the environmental Kuznets Curve, World Development 32(8), 1419-1439.

Turner, G. M., 2008. A comparison of Limimits to Growth with 30 years of reality. Global Environmental Change 18, 397 – 411.

USDA (US Dept. of Agriculture), http://usda.mannlib.cornell.edu/MannUsda/viewDocumentInfo.do?documentID=1282

USGS (US Geological Survey), http://minerals.usgs.gov/ds/2005/140/

World Bank, http://www-wds.worldbank.org/

World Bank (2012), http://data.worldbank.org/indicator/NY.GDP.MKTP.CD/countries/1W?display=default - http://www.bp.com/statisticalreview (retrieved April 2012)

York, R., Rosa, E.A., and Dietz, T., 2003. STIRPAT, IPAT and ImPACT: Analytic tools for unpacking the driving forces of environmental impacts. Environmental Economics 46, 351-365.25